\documentstyle[prl,aps]{revtex}

\draft

\input epsf         % defines \epsfbox and supporting macros

\begin{document}
%\draft command makes pacs numbers print

\twocolumn


\title{The zero kinetic energy electron effect in cold Rydberg gases}

%\title{High-Angular-Momentum Rydberg States in Cold Rydberg Gases}

%\title{$l$-Mixing, High-Angular-Momentum States
%and thermal ionization in Cold Rydberg Gases}

\author{ S. K. Dutta, D. Feldbaum, G. Raithel}

\address{University of Michigan, Physics 
Department, Ann Arbor, MI 48109-1120}

\date{\today}
\maketitle

\begin{abstract}
Cold, dense Rydberg  
gases produced in a cold-atom trap 
are investigated using spectroscopic 
methods and time-resolved electron counting. 
On the discrete Ryd\-berg resonances 
we observe large trap losses and
long lasting electron emission 
from the Ryd\-berg gas ($>$30ms).
Our observations are explained by 
quasi-elastic  $l$-mixing collisions between
Ryd\-berg atoms and slow electrons that
lead to the population of long-lived 
high-angular-momentum Ryd\-berg states. 
These atoms thermally ionize slowly and with 
large probabilities, leading to the observed effects. 
\end{abstract}


Laser-cooled atoms can be used to 
study highly excited Ryd\-berg atoms \cite{Gal} 
at both large densities and low atomic velocities.
Due to the low velocity of the Ryd\-berg atoms, 
ionizing Ryd\-berg-Ryd\-berg collisions that dominate the 
behavior of hot Ryd\-berg gases \cite{ionpair,assion,Vit82} 
are largely suppressed. Therefore, the interactions between the
Ryd\-berg atoms, free electrons and ions result in 
a variety of novel phenomena that are specific to
translationally cold Ryd\-berg gases. 
Density-dependent effects have been observed
in the resonant excitation transfer
between cold Ryd\-berg atoms \cite{Gal98,Pillet98}.
A minute increase of the frequency of the Ryd\-berg excitation laser 
leads to the production of metastable cold plasmas rather than 
cold Ryd\-berg gases \cite{Rol99}.
At a critical density corresponding to about one atom per 
atomic volume, Ryd\-berg gases might undergo a Mott 
transition that would lead to a new kind of 
metastable matter \cite{mott,RM,discharges}. 
The formation of metastable Ryd\-berg matter might 
proceed through an intermediate phase, in which the Ryd\-berg population 
accumulates in long-lived high-angular-momentum 
(high-$l$) states \cite{discharges}.
In the present paper, we show that in cold
Ryd\-berg gases high-$l$ Ryd\-berg states 
are efficiently produced by a robust mechanism,
which is, in a similar form, also at work 
in ZEKE (ZEro Kinetic Energy) electron - spectroscopy, 
a powerful technique 
that has revolutionized molecular spectroscopy \cite{ZEKE}.

\begin{figure}  [h]
\centerline{\ \epsfxsize=3.0in \epsfbox{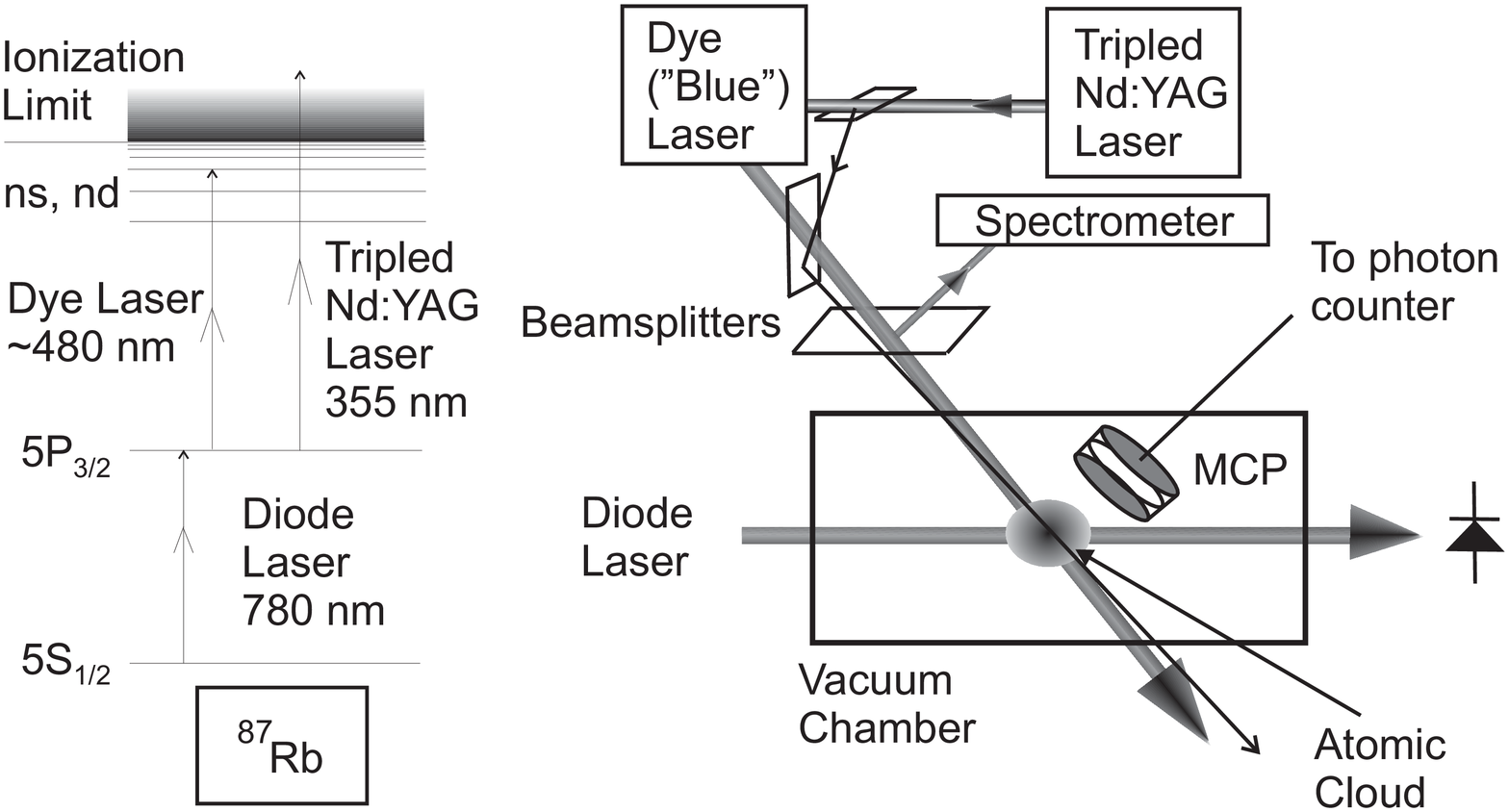}}
\caption
{Level scheme (left) and outline of the
experimental setup (right). }
\label{setup}
\end{figure}

In our periodically cycled experiment, $^{87}$Rb 
atoms are collected and cooled
in a magneto-optic trap (MOT \cite{MOT}, see Fig.~\ref{setup})  
for a time that could be varied between 
$95$ms and $950$ms. 1ms after the shutdown of the MOT   
a $5\mu s$ long diode laser pulse ($\lambda =
780$nm) resonant with the  $5S_{1/2}, F=2 \rightarrow 5P_{3/2},
F=3 $ transition is applied. While the $780$nm pulse is on, 
a blue dye laser pulse ($\lambda \approx
480$nm, 10ns width, bandwidth $\approx 15$GHz,
repetition rate 1Hz to 10Hz) excites
$ns$- and $nd$-Ryd\-berg states from the intermediate 
$5P_{3/2}$-level. The maximum photon fluence of one blue pulse easily
exceeds the saturation fluence at the photoionization threshold  
($7 \times 10^{16}$cm$^{-2}$
at $\lambda_{\rm{ion}} = 479.1$nm \cite{satflux}). 
If the dye laser operates with the oscillator only,  
the broad-band ASE (amplified spontaneous emission)
contained in the blue pulse is $< 1\%$ of the pulse energy 
($\approx 50\mu$J).
With the dye amplifiers on,  the ASE contains 
$\approx 10\%$ of the pulse energy, is 
about 5nm wide and centered at $\lambda = 478$nm, 
which is above the ionization threshold.
The pulsed dye laser is pumped 
by the 3rd harmonic of a Nd-YAG laser ($355$nm), 
a small fraction of which ($<$1mJ)
can be diverted to partially ionize 
the atomic cloud before the blue 
laser pulse arrives. 
The Ryd\-berg excitation causes a reduction of 
the ground-state population, 
which reduces the area density of 
ground-state atoms that we measure with 
a low-intensity probe laser pulse resonant on the  
$5S_{1/2}, F=2 \rightarrow 5P_{3/2}, F=3$ transition. 
A microchannel-plate (MCP) detector 
located about $10$cm from the atomic cloud is 
used to detect electrons emitted from the Ryd\-berg gas.

\vspace{-0.5cm}

\begin{figure}  [h]
\centerline{\ \epsfxsize=2.5in \epsfbox{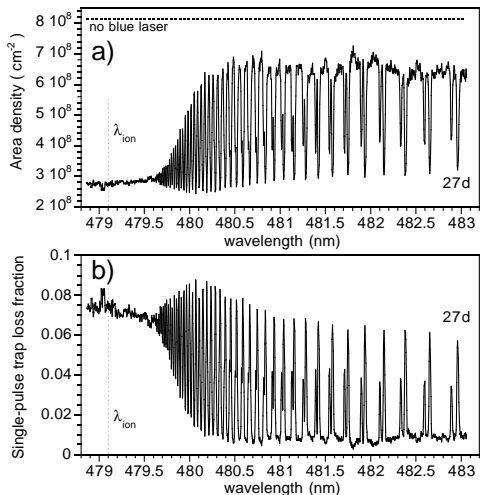}}
\caption
{Ryd\-berg excitation spectrum of atoms in
a magneto-optic trap obtained by measurements of the
trap loss at different wavelengths. Panel a): central area 
density of trapped atoms vs. the
wavelength of the blue laser. Panel b): Corresponding
estimated single-pulse loss fraction $X$.}
\label{spectrum}
\end{figure}

Fig.~\ref{spectrum}a) shows a trap-loss
spectrum taken with a 10Hz repetition rate of the 
Rydberg excitation and with both the
$5S_{1/2} \rightarrow 5P_{3/2}$ and the Ryd\-berg transition
saturated. With the blue laser blocked, 
from one cycle to the next most atoms 
become recaptured by the MOT, and the
average time the atoms remain trapped
is $2.5$s. Thus, when the blue laser is unblocked, 
the atoms experience of order
25 excitation pulses before they leave the 
trap due to the loss mechanisms that are inherent to the MOT.  
In the continuum as well as on the discrete Ryd\-berg resonances 
up to about $70\%$ of the atoms are removed by the Rydberg 
atom excitation. The trap-loss spectra are largely independent of 
the time at which the probe pulse is applied, even if the probe
is applied during the lifetime of the excited Rydberg states.
This shows that most of the observed trap loss is due to 
a {\sl permanent} removal of atoms from the trap 
upon excitation. The removal is due to many blue pulses 
exciting the same atoms. We have used a simple model of the
MOT loading dynamics to estimate 
the loss fraction $X$ due to a single Rydberg excitation event.
In Fig.~\ref{spectrum}b), a floor of $X \approx 1\% $ is 
observed between the discrete Ryd\-berg lines, 
which is due to direct photoionization by the
above discussed $10 \%$ ASE of the blue pulse.
The discrete Ryd\-berg lines peak at an $X$ of up to $8\%$
above the floor, a value that corresponds to
one-third loss of the excited Ryd\-berg atom
population. Further, there is no
significant change in $X$ at the continuum threshold.
Our observations show that there is a process by which
initially bound Ryd\-berg atoms permanently leave the trap with
an efficiency that rivals direct optical photoionization.

Two-body ionizing Ryd\-berg-Ryd\-berg collisions could, in principle,
cause a trap loss.  Based on 
an ionization cross section given in \cite{Vit82}, a Ryd\-berg atom 
velocity of $0.1$m/s, a lifetime of $100 \mu$s ($n \approx 40$),
and a Ryd\-berg atom density of $5 \times 10^{9}$cm$^{-3}$
we estimate a collisional ionization probability of $\approx 0.5\%$
of the excited Ryd\-berg population, corresponding to an $X \approx 0.13 \%$.
This figure, obtained for $n=40$, 
is too small for two-body ionizing Ryd\-berg-Ryd\-berg collisions 
to be the dominant source of trap loss.

\begin{figure}  [h]
\centerline{\ \epsfxsize=3.0in \epsfbox{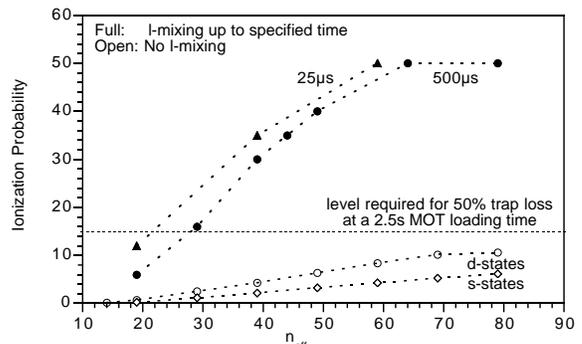}}
\caption
{Thermal ionization probabilities of Ryd\-berg
atoms vs. the effective principal quantum number
obtained from rate-equation calculations.
Open circles: $nd$-states, open diamonds: $ns$-states,
full circles: $l$-mixing for $t<500 \mu s$,
full boxes: $l$-mixing for $t<25 \mu s$. The curves are 
explained in the text.}
\label{mwion}
\end{figure}

To estimate the importance of thermally induced microwave 
ionization of the excited $ns$ and $nd$ 
Ryd\-berg states, we have performed rate equation simulations
of the population flow 
among the bound atomic states and the continuum. We use a 
basis $(n,l)$ of discrete levels up to 
$n=100$ with all allowed values of $l$ and the
proper quantum defects, and a grid of 
10100 continuum states $(\epsilon, l)$ 
with energies up to $\epsilon = 130$meV and $l=0,1,..,101$.
Due to isotropy, we can assume uniform
distributions over the magnetic substates and use 
$m$-averaged transition rates \cite{Gal}.
The obtained thermal ionization probabilities for
an ideal 300K blackbody spectrum are displayed in
Fig.~\ref{mwion}. 

While direct thermal ionization of
the initially excited states does not cause 
enough ionization to 
explain the observed trap loss - see Fig.~\ref{mwion} - it produces
ions moving at about 1m/s and electrons 
with about $8$meV average kinetic energy;
the latter figure results from the rate-equation calculations.  
Based on Fig.~\ref{mwion} and on the number of excited Ryd\-berg atoms,
we estimate that up to $\sim 10^{5}$ electron-ion pairs are created.
If the blue laser is used with its amplifiers active, 
the ionizing ASE of the blue laser pulse adds 
a significant amount of additional electron-ion pairs 
(electron energy $\approx 10$meV).
As a result, conditions are such that a 
metastable cold plasma is formed \cite{Rol99}:
a fraction of the electrons quickly evaporate, 
leaving behind a plasma with a net positive charge
that acts as an electron trap. 
If, under our conditions, the initial number of electrons exceeds
about 1000, the trap becomes
deep enough to retain a fraction of the electrons \cite{Rol99}. 
The net positive charge causes a slow Coulomb expansion,
due to which all trapped electrons eventually evaporate. 
The electron storage time is thereby limited to of order $100 \mu$s,
which is long enough for the retained 
electrons to frequently collide with the  main product 
of the laser excitation - the
bound Ryd\-berg atoms floating in the plasma. 
The collisions initiate a sequence of events 
we refer to as the ZEKE-effect.

\begin{figure}  [h]
\centerline{\ \epsfxsize=3.0in \epsfbox{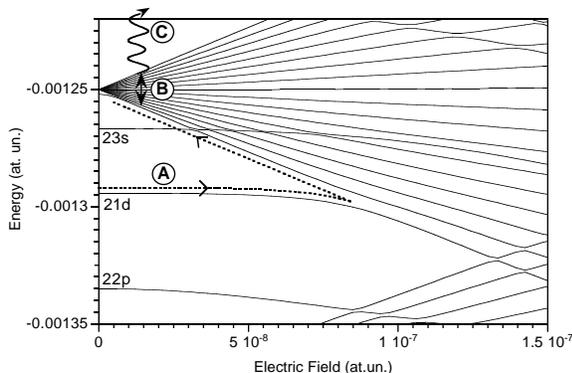}}
\caption
{The ZEKE effect visualized using 
the Stark map of Rb in the
vicinity of $n=20$. The three steps $A$, $B$ and $C$,
explained in the text, lead to the 
production of long-lived Ryd\-berg atoms and 
time-delayed thermal ionization. The  step $A$ 
is induced by quasi-elastic collisions between 
electrons and Ryd\-berg atoms.}
\label{ZEKE}
\end{figure}

In step $A$ of the 
ZEKE-effect (see Fig.~\ref{ZEKE}), the atoms are promoted from 
their initial $s$ or $d$-states 
into a high-$l$-state by the electric-field 
sweep produced by a bypassing electron.
The field sweep brings the initial Ryd\-berg 
state in contact with a hydrogenic
manifold of high-$l$ states, and state-mixing 
causes a quasi-elastic transition of the atom
into a superposition of high-$l$ states.
To model these collisions, we have numerically solved the 
time-dependent Schr\"odinger equation. 
For a given initial state $|n_0,l_0,m_{0} \rangle$, 
electron velocity $v$ and collision parameter $b$ 
the calculation yields a final probability $P(n_0,l_0,v,b)$ of 
finding the atom in the hydrogenic manifold, i.e. 
in a state with $l \ge 4$.
For $l_0 \ne 0$, we run the calculation 
for the allowed values of $m_0$ and average
the resultant probabilities over $m_0$.
The cross section $\sigma$ for a transition into $l\ge 4$ is  
then defined as 

\begin{equation}
\sigma (n_0,l_0,v) = \int_{b=0}^{b=\infty} P(n_0,l_0,v,b) 2 \pi b db
\label{sig}
\end{equation}  

For $l_0=0$, 1 or 2 it is easy to determine an upper cutoff value
for $b$ where $P \rightarrow 0$. 
The range $b < a_0 n^2$, where the neglected
ionizing processes should become dominating \cite{Vit82}, 
does not significantly contribute to $\sigma$.

\begin{figure}  [h]
\centerline{\ \epsfxsize=3.0in \epsfbox{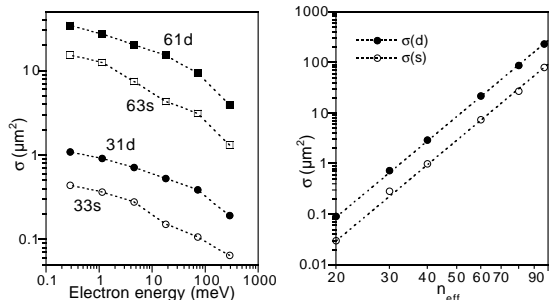}}
\caption
{Left: Calculated $l$-mixing cross sections $\sigma$ 
for collisions of Ryd\-berg atoms in the indicated 
initial states with electrons versus the 
electron energy. Right: $l$-mixing cross sections $\sigma$
with $n_{\rm{eff}}^{5}$-fits (dotted) 
for $s$- and $d$ initial states at an electron energy of 4.5meV versus the 
effective  quantum number.}
\label{cross}
\end{figure}

Fig.~\ref{cross} shows the results of our cross section
calculations for the $s$ and $d$-states of two values of $n$.  
The drop of $\sigma$ at higher electron energy 
reflects the fact that the passage behavior of the Ryd\-berg atoms
in the Stark map (Fig.~\ref{ZEKE}) turns
diabatic. The $s$-states generally 
have cross sections about 2.5 times smaller than 
the cross sections of the nearest $d$-state. This reflects the
diabatic crossing behavior of the $s$-state atoms
with the lowest few states of
the nearest hydrogenic manifold - note the narrow
anticrossings in Fig.~\ref{ZEKE}. For 
fixed electron energy, the cross sections approximately 
scale as $n_{\rm {eff} }^5$.
The average values of $l$ of the atoms that
made a transition into the hydrogenic manifold are of order $n/2$
(not shown).

The plasma volume ($\approx 1$mm$^{3}$),
the electron number ($\ge 1000$), the electron
velocity ($\approx 50000$ m/s), the electron storage time
($\approx$100$\mu$s), and the cross sections depicted in Fig.~\ref{cross}
lead to the conclusion that the step $A$ in Fig.~\ref{ZEKE} 
happens with certainty for $n$ larger than about 20.
The presence of the plasma that temporarily 
traps the electrons is crucial, as it keeps the
electrons from leaving and causes them to frequently collide
with the abundant, bound Ryd\-berg atoms floating in the plasma.
Collisions between Ryd\-berg atoms and ions are 
ineffective, because the ions are very slow
($\approx 1$m/s), as are the Ryd\-berg atoms themselves.

After step $A$ in Fig.~\ref{ZEKE}, the weak 
but rapidly varying microfields generated by 
more distant electrons will be sufficient to further
randomize the Ryd\-berg population among the
quantum-defect-free hydrogenic states (step $B$). 
Step $B$ is more probable
than step $A$, but it is not required for the
subsequent step $C$. Once all plasma electrons have 
evaporated, the only electric fields the Ryd\-berg 
atoms are still exposed to are the
fields generated by the very slowly moving ions.
Therefore, we expect that the plasma dynamics and the
internal dynamics of the Ryd\-berg atoms decouple
at about $100\mu$s after the excitation, thereby
terminating the $l$-mixing.

\begin{figure}  [h]
\centerline{\ \epsfxsize=2.5in \epsfbox{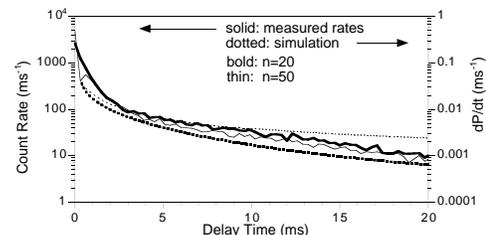}}
\caption{
Rate of detected electrons 
formed at different delay times between the Ryd\-berg
excitation and the counting gate for the indicated
Ryd\-berg $d$-states. The signal lasts at least 100 times as long as the 
natural lifetime of the initially excited states,
and shares the qualitative features of simulated results (dotted),
which show the rates at which a single Rydberg atom would 
produce a thermal electron. Note that the plasma that 
initiates the ZEKE effect decays within
$\approx 100\mu$s, and does therefore not contribute to the displayed data.
}
\label{tspec}
\end{figure}

Subsequently, while the cloud of 
high-$l$ Ryd\-berg atoms produced by 
the steps $A$ and $B$ slowly expands, the atoms 
decay or thermally ionize 
on a slow time scale
(step $C$ in Fig.~\ref{ZEKE}). To model the overall 
dynamics of the Ryd\-berg population, 
we have included an initial "plasma phase" in our 
rate-equation simulations.
During that phase, the populations
in the $n$-manifolds and the nearby non-hydrogenic
states are periodically averaged over 
the allowed $l$-values (with weights $\sim(2l+1)$);
after the ``plasma phase'' the averaging ceases.
The obtained ionization probabilities
are displayed in Fig.~\ref{mwion} for $25 \mu$s and $500 \mu$s 
long plasma phases. The
ionization probabilities are stable against variations
of the duration of the plasma phase and are large enough to 
explain the experimentally observed trap losses.

While in molecular spectroscopy the ZEKE-effect 
typically occurs for $n > 100$ \cite{ZEKE}, we observe
it down to low $n$. We believe that the large velocity  
of the charges that collide with the Rydberg atoms
- $\approx 50000$ m/s in our
experiments - and our high degree of saturation of 
the $l$-mixing process explain this difference.

The discussed model is supported by the
results described in the following.
Using the MCP detector located near the atom trap,
we have measured the thermal ionization rate  vs. 
time (Fig.~\ref{tspec}). When we excite 
discrete Ryd\-berg levels, we find 
long-lived electron signals that extend beyond $30$ms 
and that only occur if the frequency of the blue laser 
is resonant with a discrete Ryd\-berg line. The delayed 
electron signal, which is characteristic for 
the ZEKE-effect  \cite{ZEKE},
is due to thermal ionization of high-$l$ Ryd\-berg states.

\begin{figure}  [h]
\centerline{\ \epsfxsize=3.5in \epsfbox{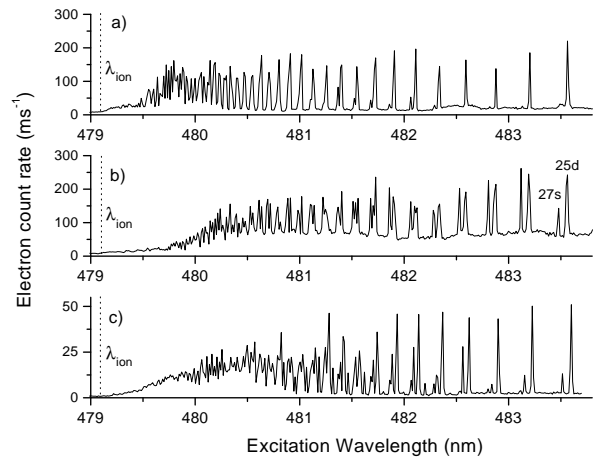}}
\caption
{Electron count rate emitted by the 
Ryd\-berg gas in a time
window from 5ms to 6ms after the excitation 
as a function of the wavelength of the blue laser. a)  
$<1\%$ ASE in the laser spectrum. b) $\approx 10 \%$ ASE
in the laser spectrum. c)  as a), but a weak 
UV pulse is used to ionize about $0.1\%$ of the atoms
a few ns before the blue laser pulse. }
\label{wspec}
\end{figure}

Fig.~\ref{wspec} shows  the electron count rate
in a time-delayed counting window  vs. the
wavelength of the blue laser for conditions
well below the saturation
of the Ryd\-berg transition. The clarity of the ionization threshold 
in the long-lived electron signal,
which is typical for the ZEKE-effect \cite{ZEKE},
shows that the delayed electrons are linked to the initial optical 
excitation of bound Ryd\-berg atoms.
In Fig.~\ref{wspec}a) the blue pulse is generated 
with the dye laser oscillator only, which 
has no significant ASE.
Therefore, all electron-ion pairs 
that are created within about $100\mu$s 
from the Ryd\-berg excitation originate 
in thermal ionization of the excited $ns$- or $nd$-Ryd\-berg atoms.
Since the $5P \rightarrow ns$-photoexcitation cross sections
are about six times smaller than the ones
of the neighboring $d$-states, and since the thermal
ionization probability of the $s$-states is
only half that of the neighboring $d$-states
(see Fig.~\ref{mwion}), the thermal electron yield on the
$s$-lines is less than one-tenth of the yield
on the neighboring $d$-lines. Recalling that
of order 1000 slow electrons are needed 
to form the essential cold-plasma electron trap - with some 
electrons left in it - it follows that
there is a wide range of parameters
where the ZEKE-signal should occur
on the $d$-lines, but not on the neighboring $s$-lines.
Fig.~\ref{wspec}a), where the $s$-lines barely appear,
shows one such case.

We have used two methods to force a ZEKE-signal 
on the $s$-lines in cases where it would
normally not appear.  In Fig.~\ref{wspec}b)
we have used the same parameters as in a), except 
that the blue laser pulse
has been generated with a dye-laser amplifier being active.  
The pulse has then been attenuated to the same pulse energy as in a).
The ASE produced by the amplifier creates
enough slow electron-ion pairs to 
form the cold-plasma electron trap,
independent of whether the coherent part of the
laser excites $d$ or $s$ Ryd\-berg atoms. As a result,
we observe both $s$ and $d$ lines in the ZEKE-signal.   
In Fig.~\ref{wspec}c), the blue laser pulse 
is the same as in a), i.e. it has no ASE, 
but we ionize a small
fraction of the atoms by a UV laser pulse 
that hits the cloud a few ns before the blue laser pulse
(see Fig.~\ref{setup}). The electrons produced
by the UV pulse have an energy of about 1eV and
therefore all leave. The remaining electron-free potential
well captures any slow electrons that are
subsequently produced. The comparatively few 
thermal electrons produced on the $s$-lines
now don't escape, as in the case of Fig.~\ref{wspec}
a), but are trapped and trigger the ZEKE-effect
(note the $s$-lines in Fig.~\ref{wspec}c)).   

\vspace{-0.5cm}

In this paper we have shown that dense, cold Ryd\-berg 
gases in a room-temperature thermal radiation field
decay via the ZEKE-effect,
which involves the quasi-elastic collisional production 
of high-$l$ Ryd\-berg states and unusually slow 
thermal ionization.
The effect hinges on the temporary existence of
a cold plasma, which acts as a  transient
electron trap. The stability 
of the observed phenomenon makes it likely that
it represents a generic decay pattern of 
cold, dense Ryd\-berg gases. It appears likely that
the spontaneous and efficient production of high-$l$ Ryd\-berg states
could aid the formation of condensed Ryd\-berg matter 
\cite{RM,discharges}.
Since it has become apparent 
that the radiation temperature 
is one of the most important 
parameters of the system, we intend to perform future
studies in a cryogenic enclosure with
variable wall temperature.  

Support by NSF is acknowledged. 
We thank Prof. P. Bucksbaum for inspiring discussions and 
generous loaning of equipment.
Equipment from a DoE-funded project
has temporarily been used.

\end{document}